\documentclass[%
 reprint,
superscriptaddress,
 amsmath,amssymb,
 aps,
prd,
longbibliography
]{revtex4-1}

\usepackage{aas_macros}
\usepackage{graphicx}
\usepackage{dcolumn}
\usepackage{bm}
\usepackage{hyperref}

\usepackage{color}
\usepackage[mathlines]{lineno}
\usepackage{tabularx}
\usepackage{subfigure}
\usepackage[normalem]{ulem}
\usepackage[export]{adjustbox}
\usepackage{url}
\hypersetup{
    pdfnewwindow=true,    
    colorlinks=true,      
    linkcolor=blue,       
    citecolor=blue,       
    filecolor=blue,       
    urlcolor=blue         
}

\newcommand\geant{\texttt{Geant4}}
\newcommand\gsolar{\texttt{G4SOLAR}}

\newcommand\nof{\texttt{NoField}}
\newcommand\quiet{\texttt{Quiet}}
\newcommand\act{\texttt{Active}}

\usepackage{natbib}

\begin{document}
\title{Simulating gamma-ray production from cosmic rays interacting with the solar~atmosphere in the presence of coronal magnetic fields}

\author{Zhe Li}
\email{lizhe@ihep.ac.cn}
\thanks{\scriptsize \!\! \href{http://orcid.org/0000-0001-8016-2170}{orcid.org/0000-0002-7065-8452}}
\affiliation{
Institute of High Energy Physics,Beijing 100049,China}

\author{Kenny C. Y. Ng}
\email{kcyng@cuhk.edu.hk}
\thanks{\scriptsize \!\! \href{http://orcid.org/0000-0001-8016-2170}{orcid.org/0000-0001-8016-2170}}
\affiliation{Department of Physics, The Chinese University of Hong Kong,Shatin,New Territories,Hong Kong}

\author{Songzhan Chen}
\email{chensz@ihep.ac.cn}
\thanks{\scriptsize \!\! \href{http://orcid.org/0000-0001-8016-2170}{orcid.org/0000-0003-0703-1275}}
\affiliation{
Institute of High Energy Physics,Beijing 100049,China}

\author{Yuncheng Nan}
\affiliation{
Institute of High Energy Physics,Beijing 100049,China}

\author{Huihai He}
\affiliation{
Institute of High Energy Physics,Beijing 100049,China}
\affiliation{School of Physical Sciences, University of Chinese Academy of Science,Beijing 100049, China}

\date{ \today}

\begin{abstract}
Cosmic rays can interact with the solar atmosphere and produce a slew of secondary messengers, making the Sun a bright gamma-ray source in the sky. Detailed observations with Fermi-LAT have shown that these interactions must be strongly affected by solar magnetic fields in order to produce the wide range of observational features, such as high flux and hard spectrum. However, the detailed mechanisms behind these features are still a mystery. In this work, we tackle this problem by performing particle-interaction simulations in the solar atmosphere in the presence of coronal magnetic fields modeled using the potential field source surface~(PFSS) model. We find that the low-energy~($\sim$\,GeV) gamma-ray production is significantly enhanced by the coronal magnetic fields, but the enhancement decreases rapidly with energy. The enhancement is directly correlated with the production of gamma rays with large deviation angles relative to the input cosmic-ray direction. We conclude that coronal magnetic fields are essential for correctly modeling solar disk gamma rays below 10\,GeV, but above that the effect of coronal magnetic fields diminishes. Other magnetic field structures are needed to explain the high-energy disk emission.

\end{abstract}
\maketitle

\section{Introduction} \label{sec:introduction}

The Sun is a high-energy astrophysical source due to its interactions with cosmic rays: Cosmic-ray electrons inverse-Compton scatter with sunlight and produce a diffuse gamma-ray halo around the Sun~\cite{Orlando:2006zs, Moskalenko:2006ta, Orlando_2021}. Synchrotron radiations by the electron would produce a disk emission from radio to X-rays~\cite{Orlando:2022xsm}. Last but not least, cosmic-ray nuclei interact with the solar atmosphere hadronically, and produce secondary gamma rays and neutrinos~\cite{Moskalenko:1991hm, Seckel:1991ffa, Moskalenko:1991hm, Ingelman:1996mj}. The latter component is mainly emitted from the photosphere, thus is more concentrated than the inverse-Compton halo; we denote it as the solar disk emission.

The solar disk gamma-ray emission was first detected with EGRET~\cite{Thompson1997, Orlando:2008uk} and later with Fermi-LAT with much better precision~\cite{Abdo:2011xn}. The observed emission is higher than early estimates by almost an order of magnitude~\cite{Seckel:1991ffa}.  Subsequent analyses with six years~\cite{Ng:2015gya} and nine years of Fermi data~\cite{Linden:2018exo, Tang:2018wqp} have found several new features, including: 1) The flux anticorrelates with solar activity at low energies~($\sim$\,1\,GeV) as well as at the highest detected energy~($\sim$\,100\,GeV) with a much larger correlation amplitude; 2) The flux exhibits a hard spectral index~($\sim E^{-2.2}$) and reaches up to at least 200\,GeV during solar minimum; 3) A spectral dip around $30–50$ GeV; 4) The photon distribution on the projected Sun disk~(solar gamma-ray morphology) varies strongly as a function of the solar cycle. See Ref.~\cite{Nisa:2019} for a brief overview and Ref.\cite{Linden:2020lvz} for the latest solar gamma-ray observation covering the full solar cycle. Currently, there are no theoretical model or calculation that can completely explain these observational features. 

The $>100$\,GeV gamma-ray emission, in particular, is highly variable; there were 6 photons observed in 1.4 years near the solar minimum, but none in the 7.8 years afterward~\cite{Linden:2018exo}. The solar minimum spectrum is also much harder than the cosmic-ray spectrum.  These suggest that very-high-energy observation of the Sun is another valuable avenue for probing the underlying physics. Only large ground-based air-shower array experiments, such as ARGO-YBJ, HAWC, and LHAASO can observe the Sun at TeV energies. ARGO-YBJ has provided the first set of strong constraints on sub-TeV to 10 TeV emission during the quiet Sun period from 2008--2010 \cite{Bartoli:2019xvu}. HAWC was able to provide a stronger constraint~\cite{Albert:2018vcq} using data from November 2014 to December 2017. Recently, HAWC has detected TeV gamma rays with a soft spectrum with 6 years of data~\cite{HAWC:2022khj}.
The Sun was more active in that period, therefore the high-energy spectrum is expected to be soft from Fermi observations. IceCube has performed the first dedicated search of solar atmospheric neutrinos~\cite{Aartsen:2019avh}, but the sensitivity is still above the predicted flux~\cite{Arguelles:2017eao, Edsjo:2017kjk}. 

The first detailed computation of the solar disk gamma-ray flux was performed by Seckel, Stanev, and Gaisser~\cite{Seckel:1991ffa}, who proposed that charged cosmic rays entering the atmosphere are reflected by concentrated magnetic flux tubes, and thus enhances the gamma-ray production compared to the zero-magnetic field case. 
Most subsequent calculations on gamma rays~\cite{Zhou:2016ljf, Gao:2017bfv} and neutrinos~\cite{Ingelman:1996mj,Arguelles:2017eao, Edsjo:2017kjk} have ignored magnetic fields.  In Ref.~\cite{Zhou:2016ljf}, the \emph{minimum} disk emission from the Sun limb was estimated with zero-magnetic field calculations, and in Ref.~\cite{Linden:2018exo}, the \emph{maximum} was estimated by assuming all cosmic rays are reflected on the solar surface and produce gamma rays with 100\% efficiency. Most recently, Ref.~\cite{Mazziotta:2020foa} and \cite{Hudson:2020MNRAS} have considered the effect of magnetic fields around the Sun in the context of solar gamma-ray production.
However, none of the calculations can explain the observations, such as the flux, spectral shape, time variability, etc. 

In this work, to better understand the phenomenologies behind cosmic rays interacting with the Sun, we study the production of solar disk emission using the particle simulation toolkit, \geant, together with the observation-based PFSS magnetic-field model.

\section{Simulation setup} \label{sec:simulation}

\subsection{The Geant4 Toolkit} \label{sec:geant4}

\geant\ is a software toolkit that simulates the passage of particles through matter \cite{Allison:2016lfl}. Due to its powerful functionality and modeling capability, \geant~is used in many applications, such as high-energy physics, nuclear physics, medical science, and space science.  We base our computation on version \texttt{10.3.3}. 

A typical \geant\ simulation contains many components, such as detector designation, event generator, particle definition, and physics models.
We focused on particles within the energy range between 100\,MeV and 100\,TeV, which covers the energy range of Fermi and HAWC. Following the user's guide~\footnote{\url{https://geant4-userdoc.web.cern.ch/UsersGuides/PhysicsListGuide/html/reference_PL/FTFP_BERT.html}} and recommended by Geant4 for high energy physics(HEP), we use the FTFP-BERT physics list to model both hadronic and electromagnetic interactions. Below 5\,GeV, the Bertini cascade model~\cite{Heikkinen:2003sc} is used.  With a transition between 4-5\,GeV, the Fritiof string model is used above 4\,GeV\cite{Yarba:2012}. During simulating particles propagation, we set the simulation procedure stop and kill the particles bellow 100MeV. So the main physical processes during tracking particles are hardronic decay, electron/positron bremsstrahlung and annihilation. 

Magnetic fields are included in \geant\ through a separate class. In order to propagate a particle inside a magnetic field, Geant4 solves the Lorentz force equation of motion of the particle. To calculate the track's motion in a field, Geant4 breaks up this curved path into linear chord segments. Following the \geant\ guide, we compute the particle tracks using the default Runge-Kutta method. We also adopted the Geant4’s default stepper “ClassicalRK4”, and set the chord segment as 10 metres. 

We have enabled the standard physics lists with Geant4 for hadronic and electromagnetic interactions, except synchrotron radiations.  Furthermore, solar photon itself is not included. Thus, both synchotron and inverse Compton energy loss is not included in the simulation. Generally, given the short distance scales of the problem, these processes are subdominant compared to hadronic and electromagnetic interactions in the matter itself, except perhaps at much higher energy than our problem at hand~\cite{Andersen:2011dz} or maybe at longer distance scales~\cite{Orlando:2022xsm}.

\subsection{\label{sec:code} The G4SOLAR code}
Based on the \geant~toolkit, we develop \gsolar, a program that handles particle propagation and particle interactions in the solar atmosphere. In this section, we describe the essential components of \texttt{G4SOLAR}. 

\begin{figure}[t!]
    \centering
    \includegraphics[width=3.5in,height=3.0in]{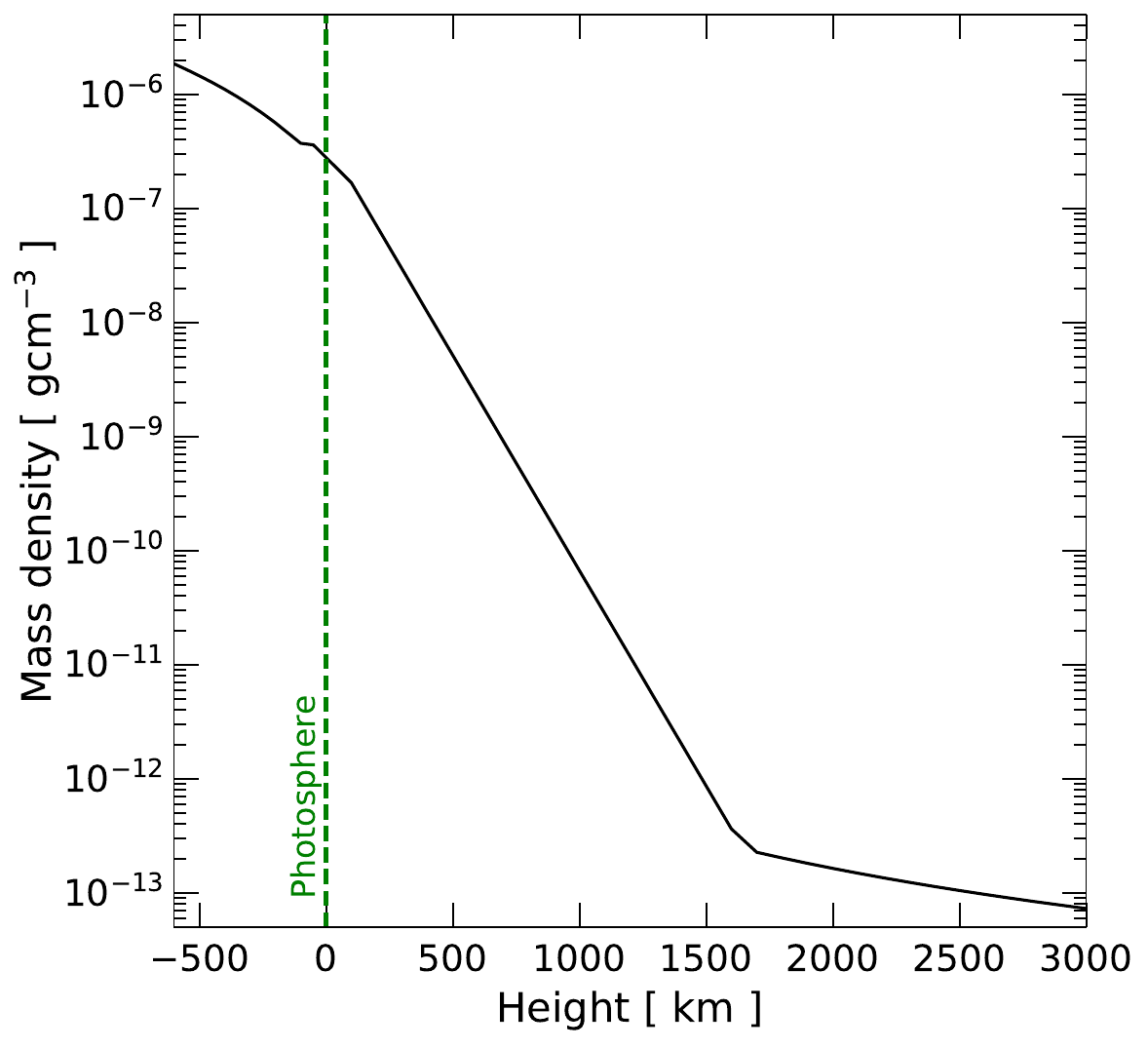}
    \caption{The density profile of the Sun near the photosphere~\cite{Seckel:1991ffa, Baker:1966, White:1966}. } 
    \label{fig:density}
\end{figure}

\subsubsection{\label{sec:density} Solar atmosphere}

The atmosphere of the Sun consists of the photosphere, the chromosphere, and the corona~\cite{Wiegelmann2014}. The photosphere, with roughly 500\,km thickness, is the layer where the Sun becomes optically opaque. The chromosphere is the region roughly a few thousand km above the photosphere; and the corona is defined as the large region above the chromosphere, where the temperature rises to millions of kelvin. 

Figure~\ref{fig:density} shows the density distribution of the solar atmosphere used in our calculation. Below the photosphere~(set at 0\,km), we use the density provide by Ref.~\cite{Baker:1966}, and we use Eq.~(2.1) of Ref.~\cite{Seckel:1991ffa} to extend it to 1600\,km.  Between 1600 to 3000\,km, we use the data in Ref.~\cite{White:1966}. We configure the Sun as a sphere and divide the region from -600\,km to +3000\,km into 3600 equal-thickness layers following the density profile shown in the figure. 

\subsubsection{Solar magnetic fields} \label{sec:PFSS}

For magnetic fields near the Sun, we consider the potential field source surface~(PFSS) model~\cite{Schatten:1969, Altschuler:1969SoPh, Hoeksema:1984PhDT, Wang:1992ApJ}, which describes the large-scale magnetic fields above the photosphere, $R_{\odot}$. Using the photospheric magnetic-field measurement as a boundary condition, and assuming that the current density is zero as well as the fields are completely radial at a distance, $R_{ss}\sim {\cal O}(1)R_{\odot}$, the magnetic fields between $R_{\odot}$ and $R_{ss}$ can be computed by solving for the scalar potential. Thus, for each complete Carrington cycle~($\sim27$ days), using the photospheric measurements by observatories such as  GONG~\footnote{\url{https://gong.nso.edu}}, and SOHO/MDI~\footnote{\url{https://sdo.gsfc.nasa.gov/}}, a PFSS model can be obtained with only $R_{ss}$ as a free parameter, which is fitted separately.  We note that the large-scale magnetic fields we consider are drastically different from the small-scale magnetic flux tube used by Seckel et al.~\cite{Seckel:1991ffa}; we compare with their results in detail in Sec.~\ref{sec:previous_work}.  The PFSS model is easy to implement and was found to agree reasonably well with more detailed and computationally expensive magnetohydrodynamic models~\cite{Riley:2006ApJ}, thus making it a natural choice for our simulation study.

\begin{figure}[t!]
    \centering
    \includegraphics[width=3.0in,height=2.5in]{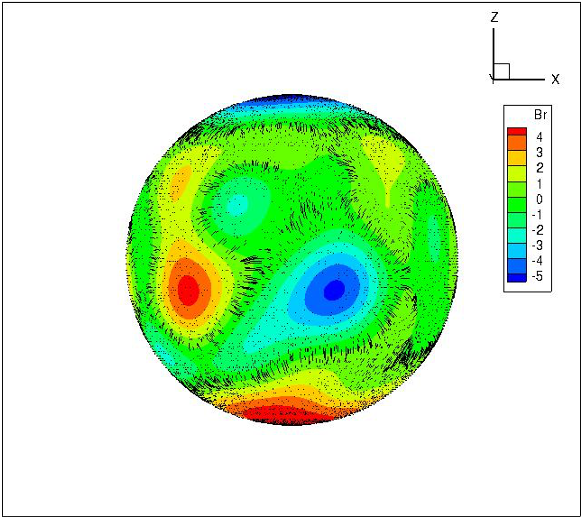}
    \includegraphics[width=3.0in,height=2.5in]{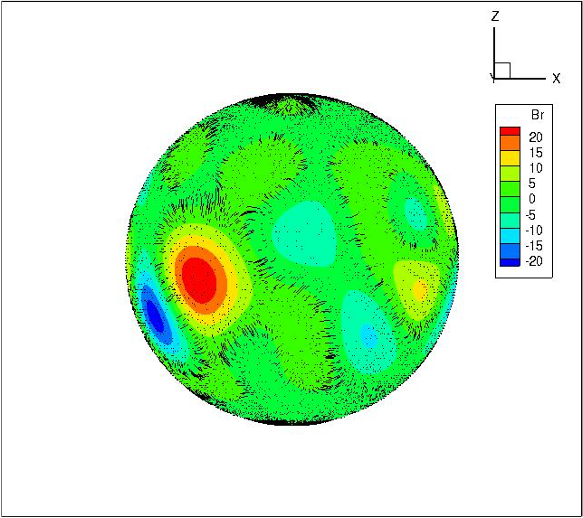}
    \caption{The obtained magnetic field distribution in spherical coordinates for both \quiet\ (above) and \act\ (bellow) concerning to CR2070 and CR2149, respectively. Color-bar for $B_r$ is in Gauss unit. Black lines indicate the magnetic field line structure at a solar radius close to photosphere surface. The dipole structure in the \quiet\ case can be clearly seen. The field strength in the \act\ case is overall stronger, and the dipole structure become unclear. }
    \label{fig:Bview}
\end{figure}

The PFSS models are obtained using the Solar Software~(SSW) package~\footnote{\url{http://www.mssl.ucl.ac.uk/surf/sswdoc/solarsoft/ssw_install_howto.html}}. We consider the Carrington Rotations 2070(CR2070)~(13 May 2008 to 9 Jun 2008) for solar minimum and 2149(CR2149)~(07 Apr 2014 to 04 May 2014) for solar maximum. In the PFSS model, we use spherical harmonic coefficients with order 9 to calculate magnetic fields. The source surface parameter is chosen to be 1.6$R_{\odot}$ for solar minimum, 2.5$R_{\odot}$ for solar maximum. 
Hereafter, we denote the solar minimum case as \quiet\, the solar maximum case as \act\, and the control case with zero magnetic field as \nof. The magnetic field we obtained by PFSS for both \quiet\ and \act\ are visualized in Fig.\ref{fig:Bview}. For the \quiet\ case, the $B_r$ is  mainly distributed in two poles, and the overall magnetic field strength is smaller than \act\ case.  For the \act\ case, Br is mainly distributed near the equator.

We evaluate the PFSS model in a 361~(radial) $\times$ 180~(polar) $\times$ 360~(azimuthal) grid in our simulation volume. This resolution means that we neglect small-scale magnetic fields variations that have size below roughly $10^{4}$\,km on tangential direction. 
We note that typically the PFSS model starts at the photosphere.  For our purpose, we extrapolate the PFSS model down to 600\,km below the photosphere by setting the fields to be the same as that at the surface. Because we start our simulation at +3000\,km, we also practically ignore all the magnetic fields above this height. During simulation, the value of the magnetic fields at each point in the simulation volume is then obtained by interpolating these grid points.
In this work, we have neglected the magnetic field variation below the resolution of the magnetic field grid points~\footnote{We thank the anonymous referee for pointing this out.}.  These resolutions are limited by computation power, and ultimately also by the PFSS model itself.  These field variations could affect the propagation of low energy particles that have the Larmor radius below the grid size (about $10^4$km in the tangential direction, and 10km in the radial direction).  We leave the modeling of sub-grid effects for future work.

\begin{table} 
\caption{The mean values for the three components of the PFSS magnetic fields and their standard deviations for \quiet\ and \act\ cases, evaluated at the photosphere. }
\label{table:PFSS}
 \begin{tabular}{p{2cm}<{\centering} p{2.2cm}<{\centering} p{2cm}<{\centering} p{2cm}<{\centering}}
 \hline
  & $<B_{r}>~[G]$ & $<B_{\theta}>~[G]$ &  $<B_{\phi}>~[G]$ \\
 \hline
 \quiet\  & 0.03  $\pm$ 2.16  & 0.11 $\pm$ 2.22 & -1.04 $\pm$ 2.45\\ 
 \act\ & -0.35 $\pm$ 15.42 & 0.66 $\pm$ 15.46 & 0.77 $\pm$ 2.54 \\
 \hline
\end{tabular}
\end{table}

\begin{figure}[t!]
    \centering
    \includegraphics[width=3.0in,height=2.5in]{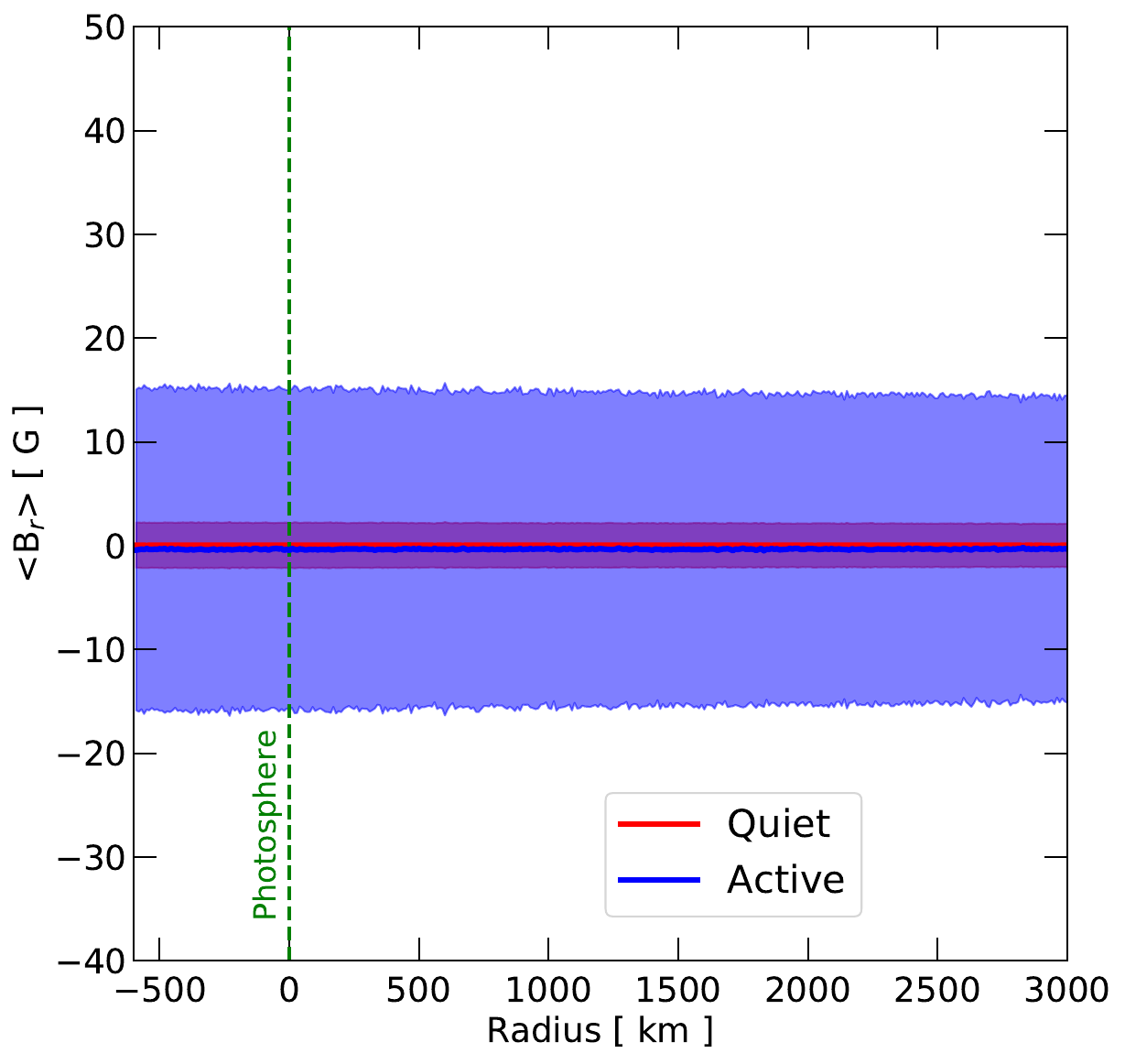}
    \includegraphics[width=3.0in,height=2.5in]{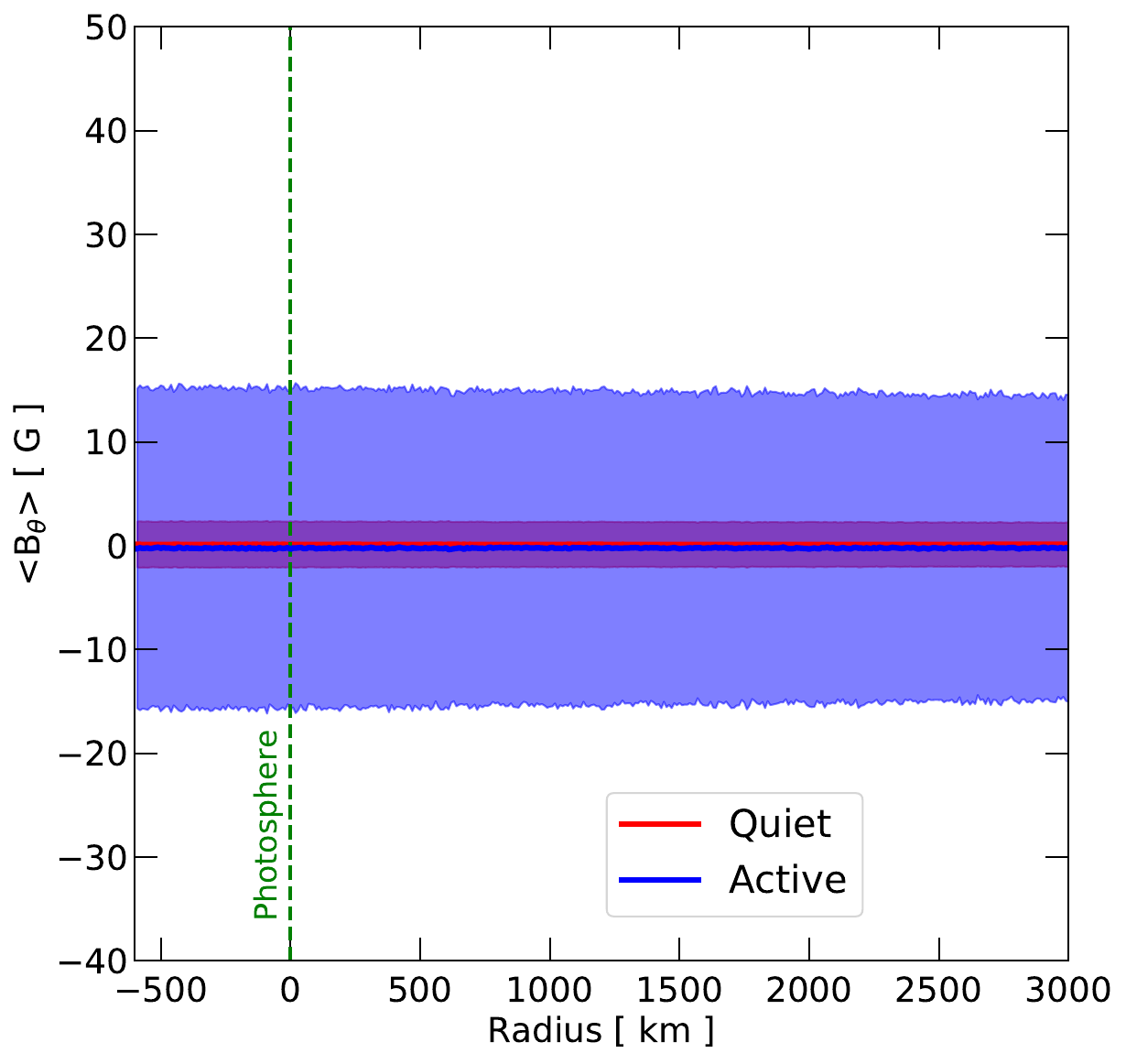}
    \includegraphics[width=3.0in,height=2.5in]{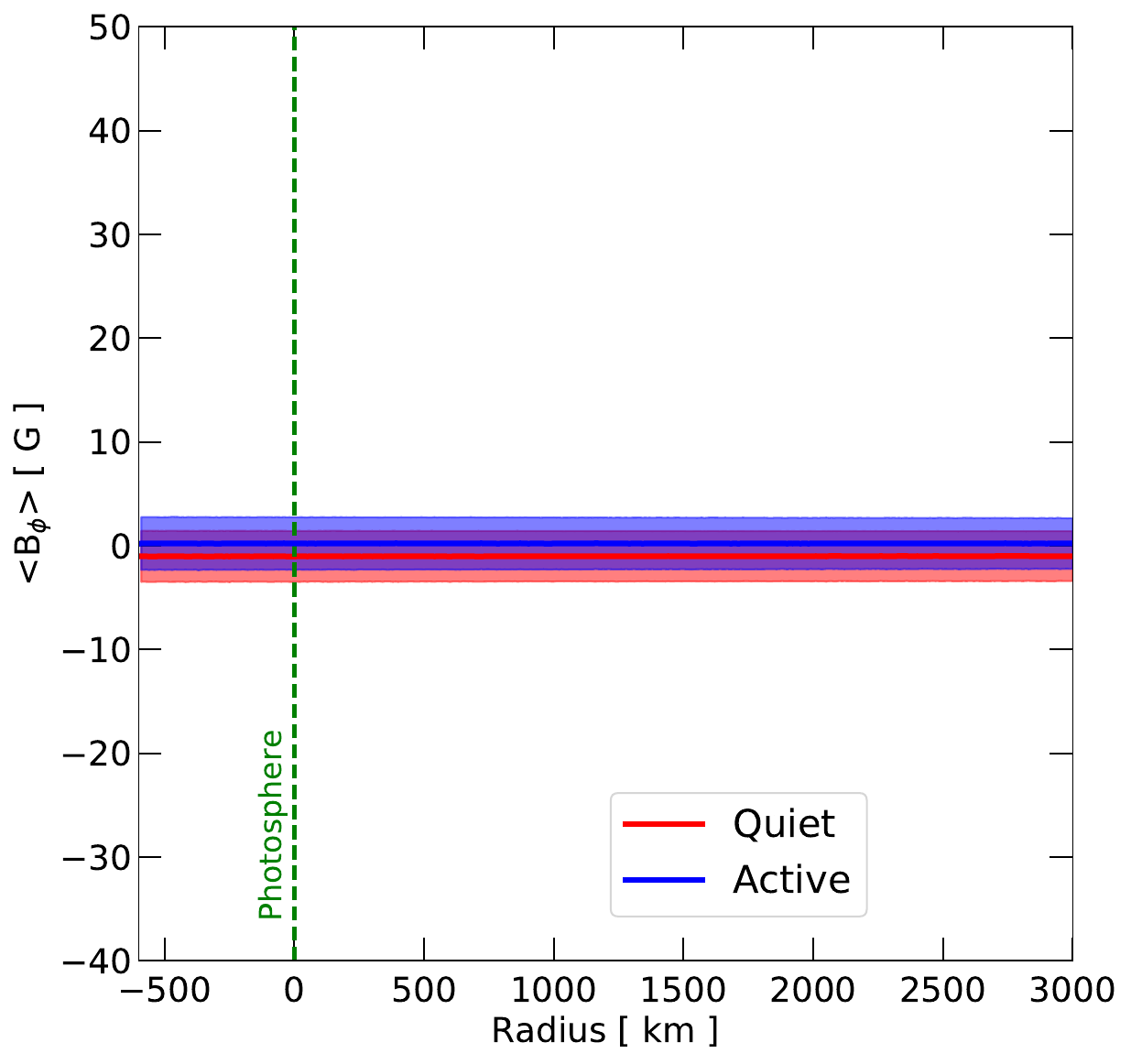}
    \caption{From top to bottom, the figures showcase the average magnetic field components B$_r$, B$_{\theta}$, B$_{\phi}$ at distances of 600 km below the photosphere and 3000 km above. Above our simulation volume at 3000 km, the field strength is set to be zero. The solid lines and shaded bands indicate the average values and deviations, respectively. "Quiet" represents the solar minimum of CR2070, and "Active" represents the solar maximum of CR2149.} 
    \label{fig:Bp}
\end{figure}

Table ~\ref{table:PFSS} and Figure~\ref{fig:Bp} show the mean values of the magnetic fields and their standard deviations for the two phases of solar activity with the PFSS model. We obtain the mean and the deviation values by sampling 100,000 points randomly at the photosphere. These values are stable versus height. Changing the sampling point between -600\,km and +3000\,km change the values by a few percent. In general, the mean values are close to zero, which is due to averaging regions with magnetic fields of opposite directions.  The standard deviation is thus more representative of the typical field strength of the model, which is $\simeq 2$\,G for the \quiet\ case. For the \act\ case, however, the standard deviation is much larger, $\simeq 15$\,G for the $r$ and $\theta$ component. 

\subsubsection{Particle sampling} \label{sec:sampling}

The final component of \texttt{G4SOLAR} is the position and direction sampling of the cosmic-ray particles. The starting position of the input particles are first sampled uniformly at 3000\,km above the photosphere. We consider the energy between 100\,MeV and 100\,TeV, which is divided uniformly into six logarithmic intervals. For each interval, the energy is sampled with a spectral index -1~(uniform in log.), and have sampling size varies due to computational time consideration, with the number of particles between $10^5$ and $10^6$. 

The momentum vectors of the particles at the input position also need to be sampled. We define the incident angle, $\omega_{p}$, as the angle between the momentum vector and the normal direction of the spherical simulation volume at particle position~(see Sec.~\ref{sec:angular}). The number of particles is then sampled according to $N \propto \sin\omega_{p} \cos\omega_{p}$. Here $\sin\omega_{p}$ is the solid angle factor and $\cos\omega_{p}$ takes into account the geometric factor between the incoming cosmic-ray flux and the receiving surface element. The azimuthal direction of the particles are sampled uniformly. In our setup, only events with $\omega_{p} \geq 90^\circ$ can enter and interact with the Sun, we thus only sample in the range between $90^{\circ}$ and $180^{\circ}$. 

All the particles are tracked only when they are in the simulation volume. Thus, for particles leaving the -600\,km layer, we assume they are completely absorbed; for particles leaving the 3000\,km layer, we assume they have escaped. The simulation results thus consist of all the escaped gamma-ray events.

\subsection{Cosmic-ray spectrum and output flux} \label{sec:flux}

To connect the simulation results to real world situations, the cosmic-ray spectrum is required. It is well-known that the Sun can change the cosmic-ray propagation environment in the solar system~\cite{Gleeson:1968zza}, and modulate the cosmic-ray flux as they propagate inward from the interstellar space. It is thus natural to expect additional modulation exist when cosmic rays propagate from Earth orbit to the vicinity of the Sun. 
However, cosmic-ray propagation in the solar system is still an open problem~\cite{Potgieter:2013pdj}, and is likely important only at low energies. For simplicity, we use the cosmic-ray spectrum measured at the Earth position as our default result. Solar modulation suppresses the low-energy cosmic rays, thus, it could suppress the actual gamma-ray production compared to our default results. We discuss and further investigate the effect of solar modulations in Sec.~\ref{sec:modulation}. 

We set the composition of the Sun to be 100\% protons, and only consider cosmic-ray protons. Including heavier species, such as Helium in both cosmic rays and the atmosphere could enhance the gamma-ray production.  We discuss this and estimate the effect in Sec.~\ref{sec:nuclei}. 

We use the 2006 proton spectrum by \texttt{PAMELA}~\cite{Adriani:2013as} from 0.1\,GeV to 45\,GeV, then \texttt{AMS-02}~\cite{Aguilar:2015ooa} from 45\,GeV to 2.5\,TeV, and finally \texttt{CREAM}~\cite{Yoon:2011aa} up to 100\,TeV.  

All the photon events leaving the simulation volume are collected and re-weighted to account for the cosmic-ray spectrum, and normalized to obtain the photon luminosity per unit area~(at the injection radius) per unit time. The photon flux at Earth is then obtained by taking into account the scaling factor $(R_{\odot} + 3000\,{\rm km})^2/{\rm AU}^2$. The $\gamma$-ray flux at the earth, $F_{\gamma,\oplus}$, are obtained by converting the simulated production yield,
\begin{equation}\label{eq:fluxcal}
    F_{\gamma,\oplus}=\frac{dN_{\gamma,mc}}{dE_{\gamma}dSdt} \cdot \frac{F_{p,\odot}}{F_{mc}} \cdot \frac{(R_{\odot} + 3000\,{\rm km})^2}{{\rm AU}^2} \, ,
\end{equation}
where $dN_{\gamma,mc}/dE_{\gamma}dSdt$ is the normalized $\gamma$-ray yield (event per energy per area per time) obtained at the $R_\odot$+3000\,km surface, $F_{p,\odot}$ is the proton spectrum from observation, $F_{mc}$ is the proton spectrum we put in the simulation.

\section{Simulation Results}\label{sec:results}

\subsection{Zero magnetic field case} \label{sec:zerofield}

We first consider the \nof\ case as a cross check and validation of the simulation procedure.  The simulation and analysis are performed without magnetic fields, otherwise keeping all procedures identical to the cases with magnetic fields. 

\begin{figure}[t!]
    \centering
    \includegraphics[width=\columnwidth]{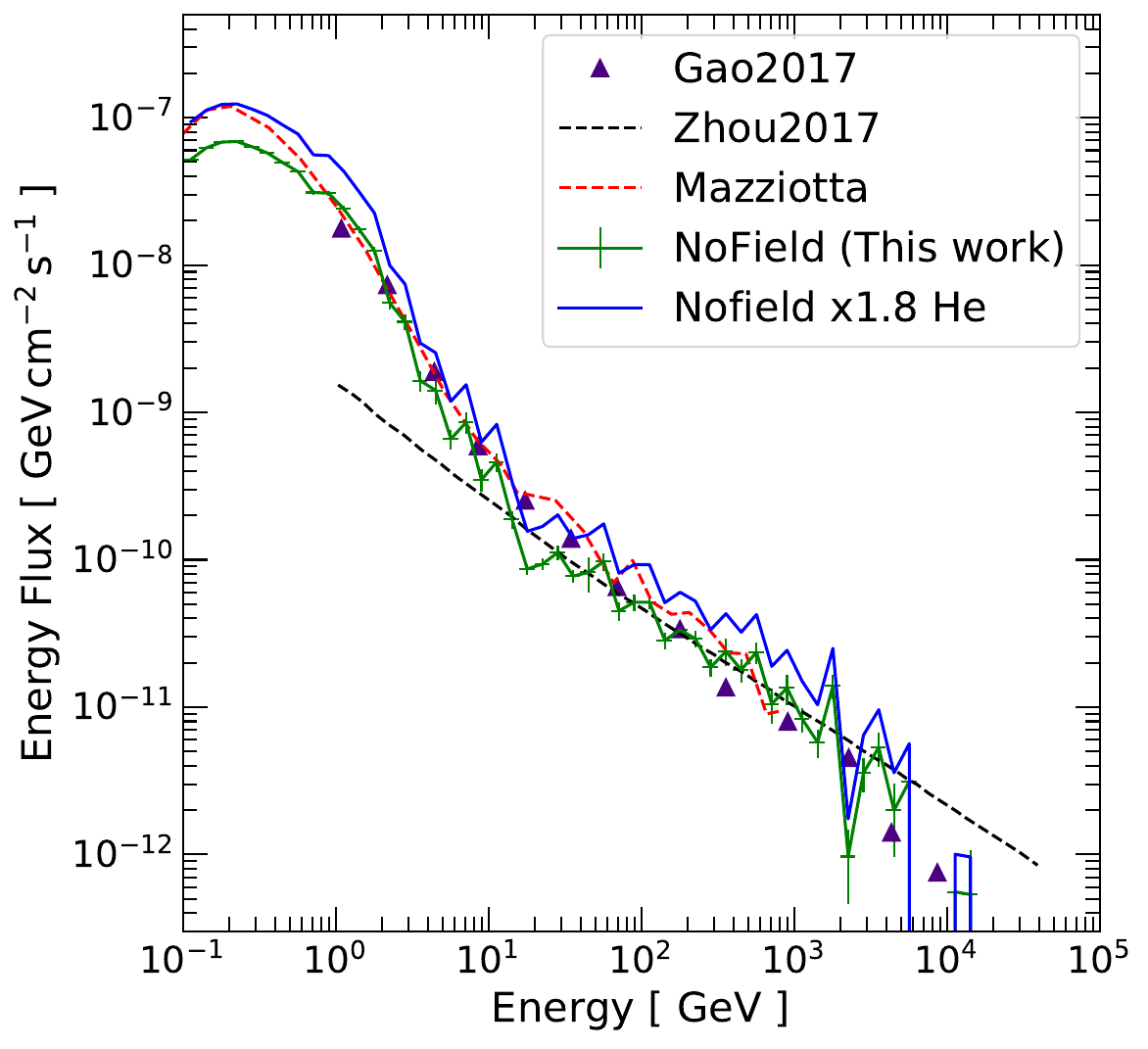}
    \caption{Simulated solar disk gamma-ray flux without magnetic fields~(\nof). For comparison, we also show the 1D semi-analytic calculation by Zhou et al.~\cite{Zhou:2016ljf}, the simplified \geant\ simulation result by Gao et al.~\cite{Gao:2017bfv}, and the simulation result with FLUKA by Mazziotta et al.~\cite{Mazziotta:2020foa}. The blue solid line is our result but scaled with the nuclei effects. ~(See Sec.~\ref{sec:nuclei}) The enhanced gamma-ray production below 10\,GeV is due to large-angle events caused by kinematics. See Sec.~\ref{sec:nofield_angle} for details.
    }\label{fig:nofield_flux}
\end{figure}

Figure~\ref{fig:nofield_flux} shows the results for the \nof\ case. We compare with the semi-analytic calculation by Zhou et al.~\cite{Zhou:2016ljf}, where they assumed that the incoming cosmic rays and the gamma rays produced are collinear. With this approximation, only cosmic rays that point at the Earth and graze through the edge of the solar atmosphere can produce detectable gamma rays~(Sun limb). This limb flux can be easily calculated as it is a 1D computation; the resultant flux roughly follows the cosmic-ray spectral index. We find that our \nof\ results agree well with Zhou et al. above 10\,GeV, meaning that the collinear approximation is appropriate here. Below 10\,GeV the \nof\ case has much higher gamma-ray production, which is caused by large-angle gamma-ray events. We discuss this in more detail in Sec.~\ref{sec:nofield_angle}. Above a few TeV, we start to see deviations due to the imposed cosmic-ray energy cutoff at 100\,TeV.  

We then compare our results with Gao et al.~\cite{Gao:2017bfv}, the precursor of this work, where the cosmic-ray position sampling step was not implemented due to the spherical symmetry of the problem when magnetic fields are not included. We find that our results agree well with each other. This validates our 3D position and angle sampling procedures described in Sec.~\ref{sec:geant4}, which are necessary once global solar magnetic fields are introduced.

We also compare our \nof\ results with that from Mazziotta et al.~\cite{Mazziotta:2020foa}, which performed essentially the same calculation but with the FLUKA simulation package. Taking into account the contribution due to nuclei effects~(see Sec.~\ref{sec:nuclei}), our results are in excellent agreement with each other, up to some fluctuations due to simulation.

\subsection{Results with magnetic fields} \label{sec:bfield_flux}

\begin{figure}[t!]
    \centering
    \includegraphics[width=\columnwidth]{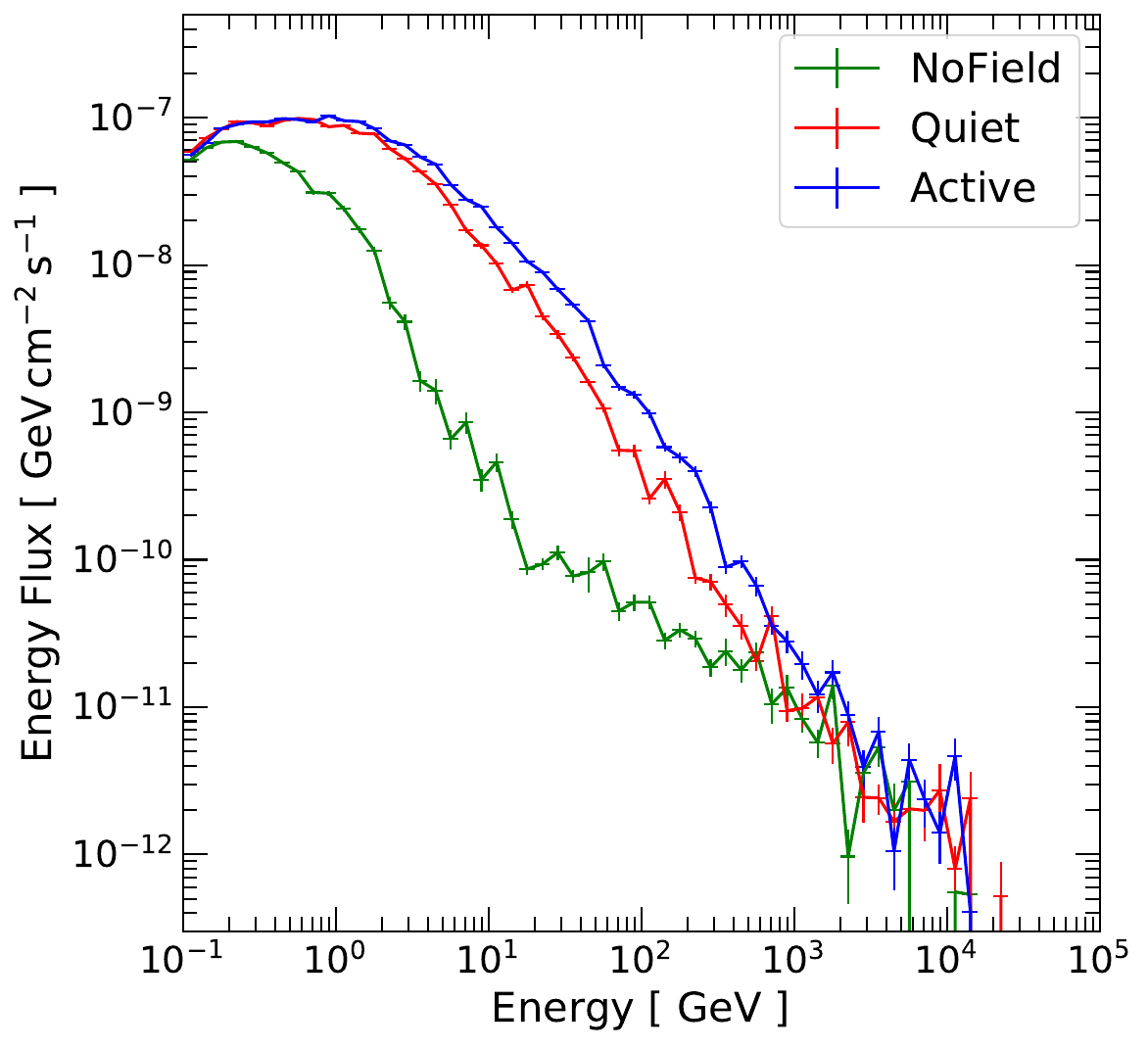}
    \caption{
    Solar disk gamma-ray flux computed by \gsolar\ without magnetic fields~(\nof), with solar minimum PFSS magnetic fields~(\quiet), and with solar maximum PFSS magnetic fields~(\act). The magnetic fields boost the gamma-ray production by enhancing the production of large-angle events. See Sec.~\ref{sec:bfield_angle} for details.
    } \label{fig:bfield_flux}
\end{figure}

Figure~\ref{fig:bfield_flux} shows the solar disk gamma-ray flux for \quiet\ and \act\ together with that for \nof. We find that the PFSS magnetic fields can dramatically change the gamma-ray production. 
At 100\,MeV, all three cases have similar flux. (See Sec.~\ref{sec:bfield_angle} for further discussion.)
Between 100\,MeV and 10\,GeV, though, both \quiet\ and \act\ exhibit harder spectral shapes and have higher flux than the \nof\ case.  The difference in flux is the largest at around 10\,GeV, by almost two orders of magnitude. Above 10\,GeV, the spectra fall sharply, and have spectral shapes even softer than the cosmic-ray spectrum. Around 1\,TeV, the \quiet\ flux and \act\ flux merge with the \nof\ flux, showing that the PFSS magnetic fields can no longer affect the gamma-ray production. We interpret these observations using the event angular distributions in Sec.~\ref{sec:bfield_angle}.

\begin{figure*}[t]
    \includegraphics[width=2\columnwidth]{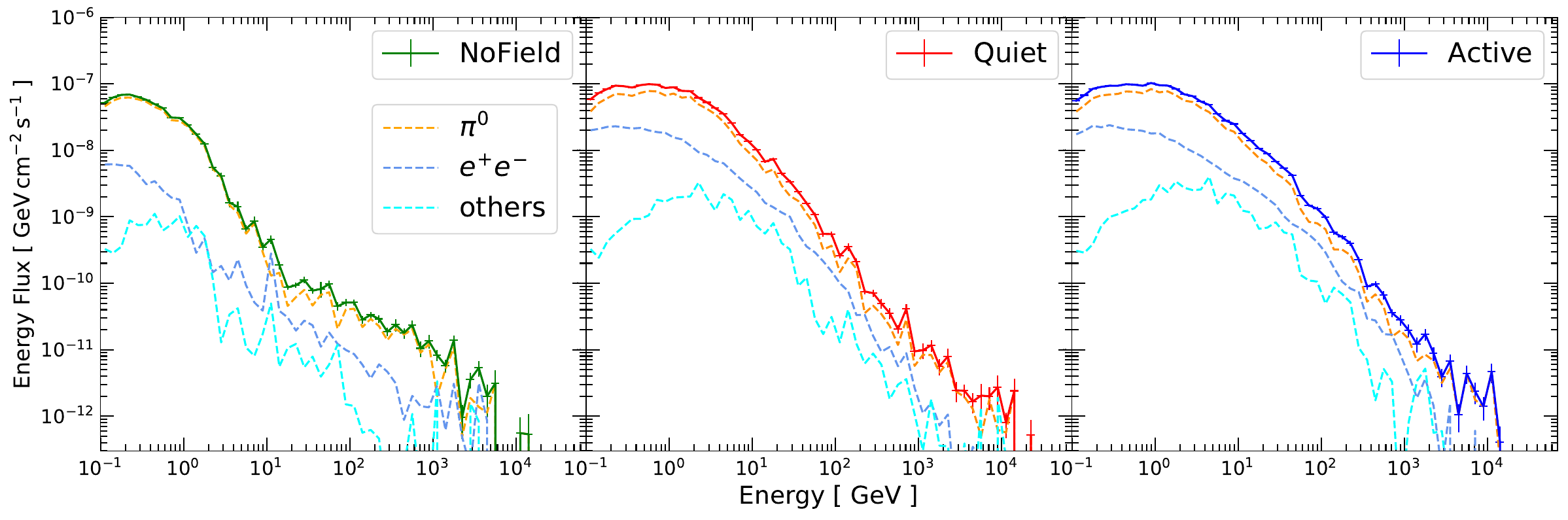}
    \caption{The breakdown of the total flux into the physical processes that are responsible for the gamma-ray production. The dominant process is neutral pion decays, then followed by electron/positron bremsstrahlung and annihilation, and finally miscellaneous processes that include decay of heavier hadron states, etc. See text for details.} 
    \label{fig:production}
\end{figure*}

\begin{figure*}
    \includegraphics[width=2\columnwidth]{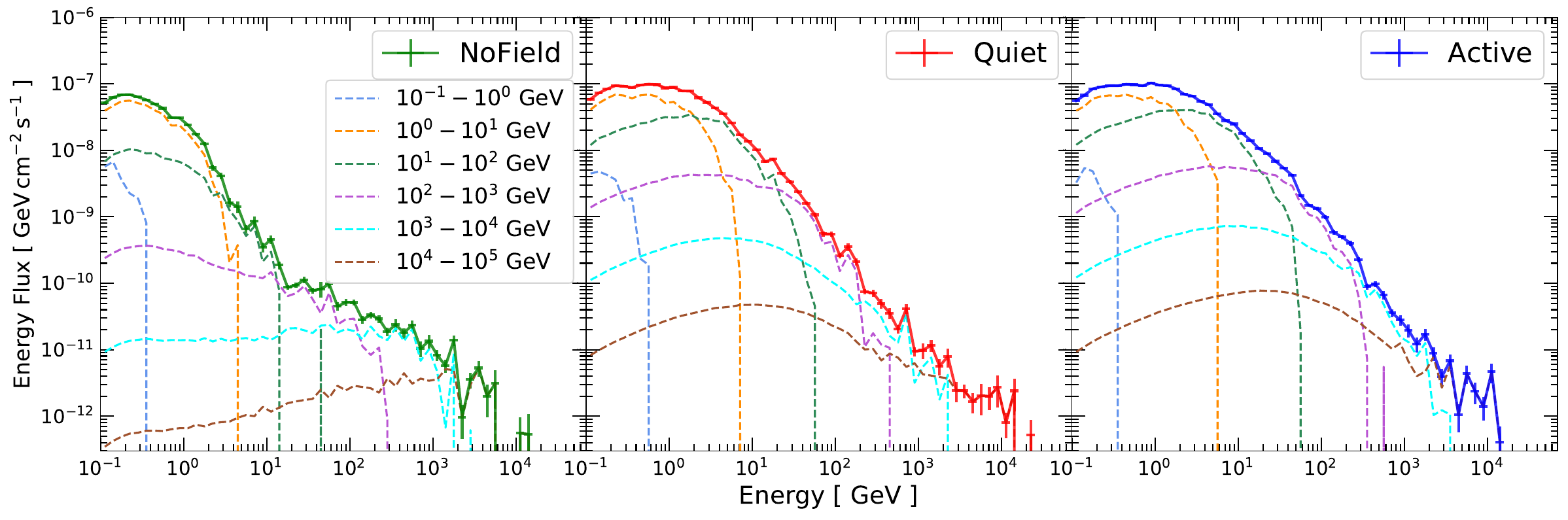}
    \caption{The breakdown of the total flux into the contributions from the corresponding input proton energy intervals. Vertical lines correspond to isolated nonzero energy bins.  } 
    \label{fig:flux_energy}
\end{figure*}

\begin{figure}[t!]
    \centering
    \includegraphics[width=\columnwidth]{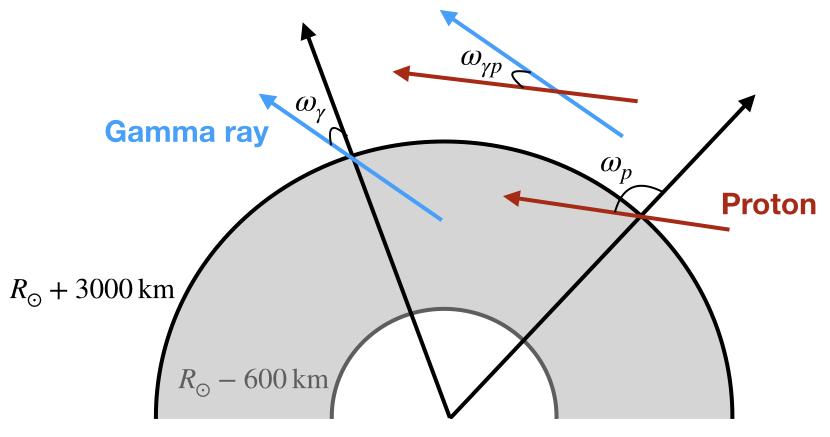}
    \caption{The definitions of the angles considered in Sec.~\ref{sec:angular}. Schematically, the grey region highlights the simulation volume. The black arrow is the normal direction from the center of the Sun and the dark red~(blue) arrow is the proton~(gamma-ray) velocity vector.  }
    \label{fig:diagram}
\end{figure}

Comparing the results between \quiet\ and \act, the two fluxes have similar shapes, except that the \act\ flux becomes larger by about a factor of two in 1\,GeV to 1\,TeV. 

In the next few subsections, we explore in detail the simulation results and attempt to understand various properties of these results.

\begin{figure*}[t!]
    \centering
    \includegraphics[width= 2.0\columnwidth]{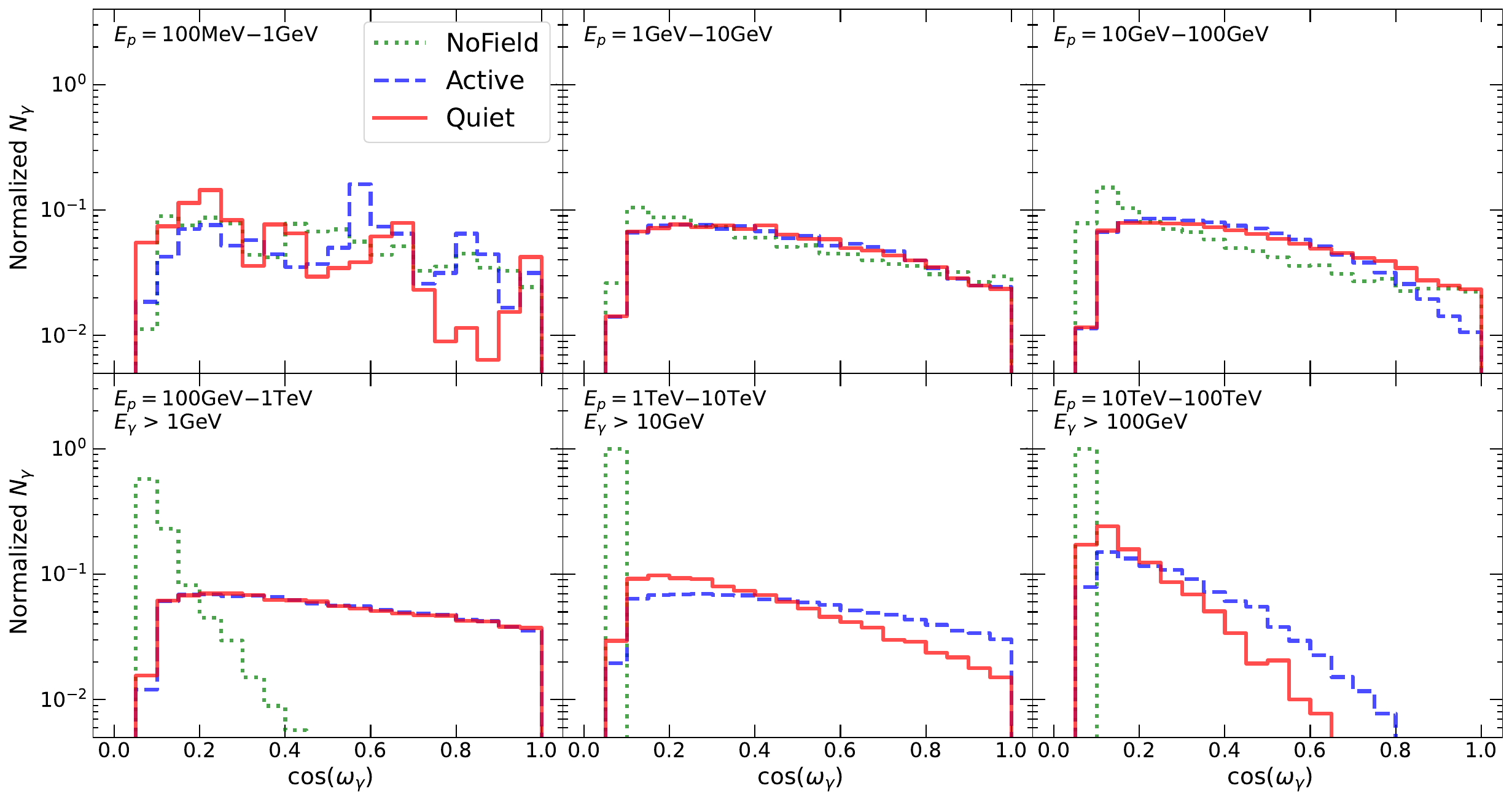}
    \caption{The angular distribution of the gamma rays produced in \nof, \quiet, and \act. $\cos(\omega_{\gamma}) = 1$ corresponds to the radially outward direction and $\cos(\omega_{\gamma}) = 0$ corresponds to the tangential direction. Each panel correspond to an interval of the input proton energies, and the distributions are all renormalized to have the same area. For display purpose, gamma-ray energy cuts are applied in the bottom three panels. 
    At high energies, all three distributions peak at $\cos(\omega_{\gamma}) \simeq 0$, which corresponds to the Sun-limb scenario~\cite{Zhou:2016ljf}. However, the cases with magnetic fields have broader distributions at all energies, especially  at medium energies.  These large-angle contributions are responsible for the enhanced solar gamma-ray production compared to the \nof\ case.  It is worth noting that at low energies, the \nof\ distribution is also broadened (with the gamma-ray flux enhanced), which is likely caused by kinematic effects.}  \label{fig:omega_gamma}
\end{figure*}

\begin{figure*}[t!]
    \centering
    \includegraphics[width= 2.0\columnwidth]{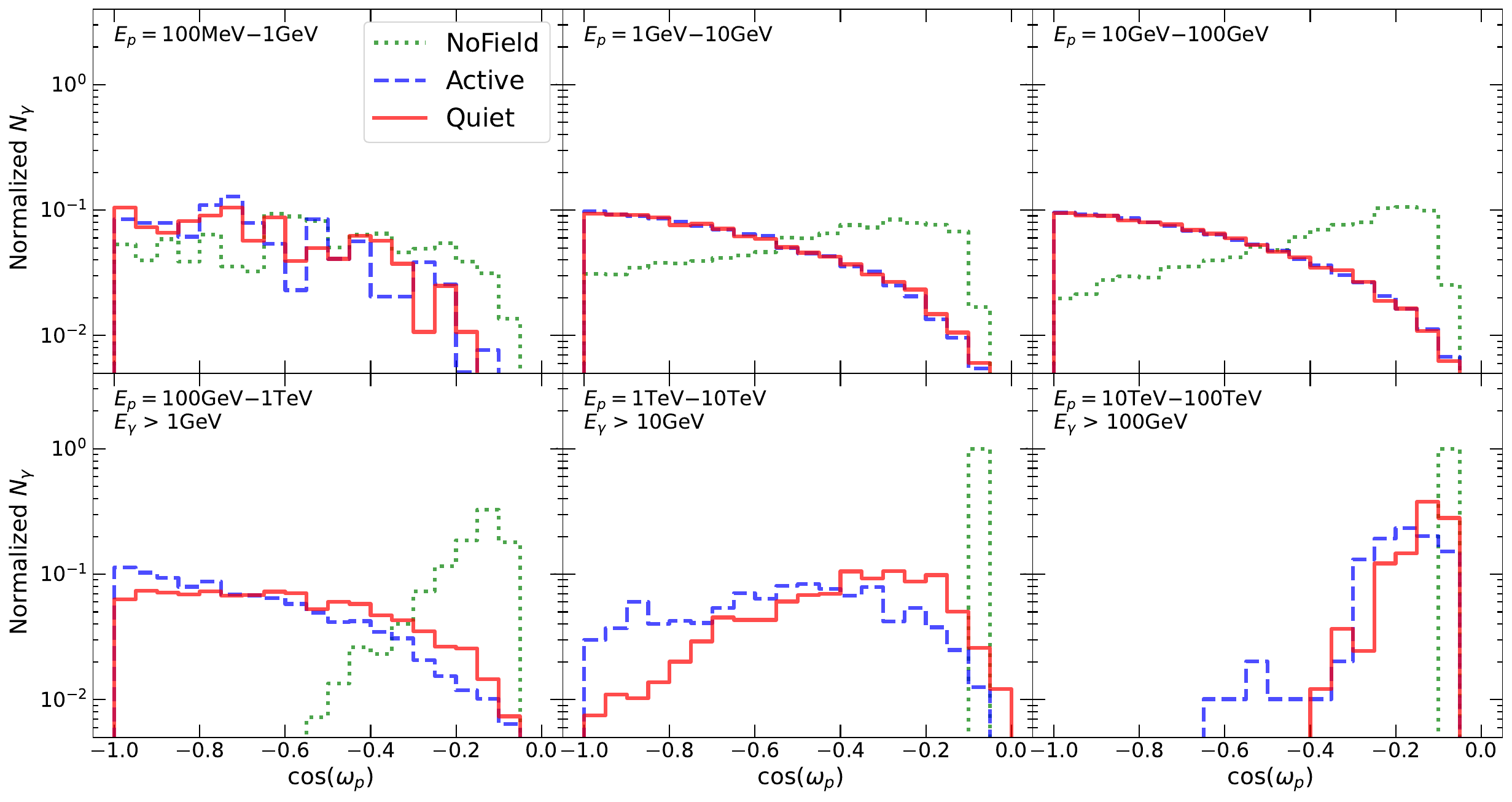}
    \caption{Similar to Fig.~\ref{fig:omega_gamma}, but for the angular distribution of the incoming protons that have successfully produced an outgoing photon.  
    $\omega_{p}$ is defined as the angle between the incoming proton vector and the normal direction of the simulation volume at the proton injection position. Therefore, $\cos{\omega_{p}} \simeq 0$ corresponds to the Sun-limb scenario. The inclusion of magnetic fields broadens the angular distributions, which are responsible for the enhanced gamma-ray production.  It is worth noting that the higher \act\ flux compared to the \quiet\ case from $E_{p} >$ 100\,GeV~(Fig.~\ref{fig:flux_energy}) is reflected here by the broader angular distributions. } 
    \label{fig:omega_proton}
\end{figure*}

\begin{figure*}
    \centering
    \includegraphics[width= 2.0\columnwidth]{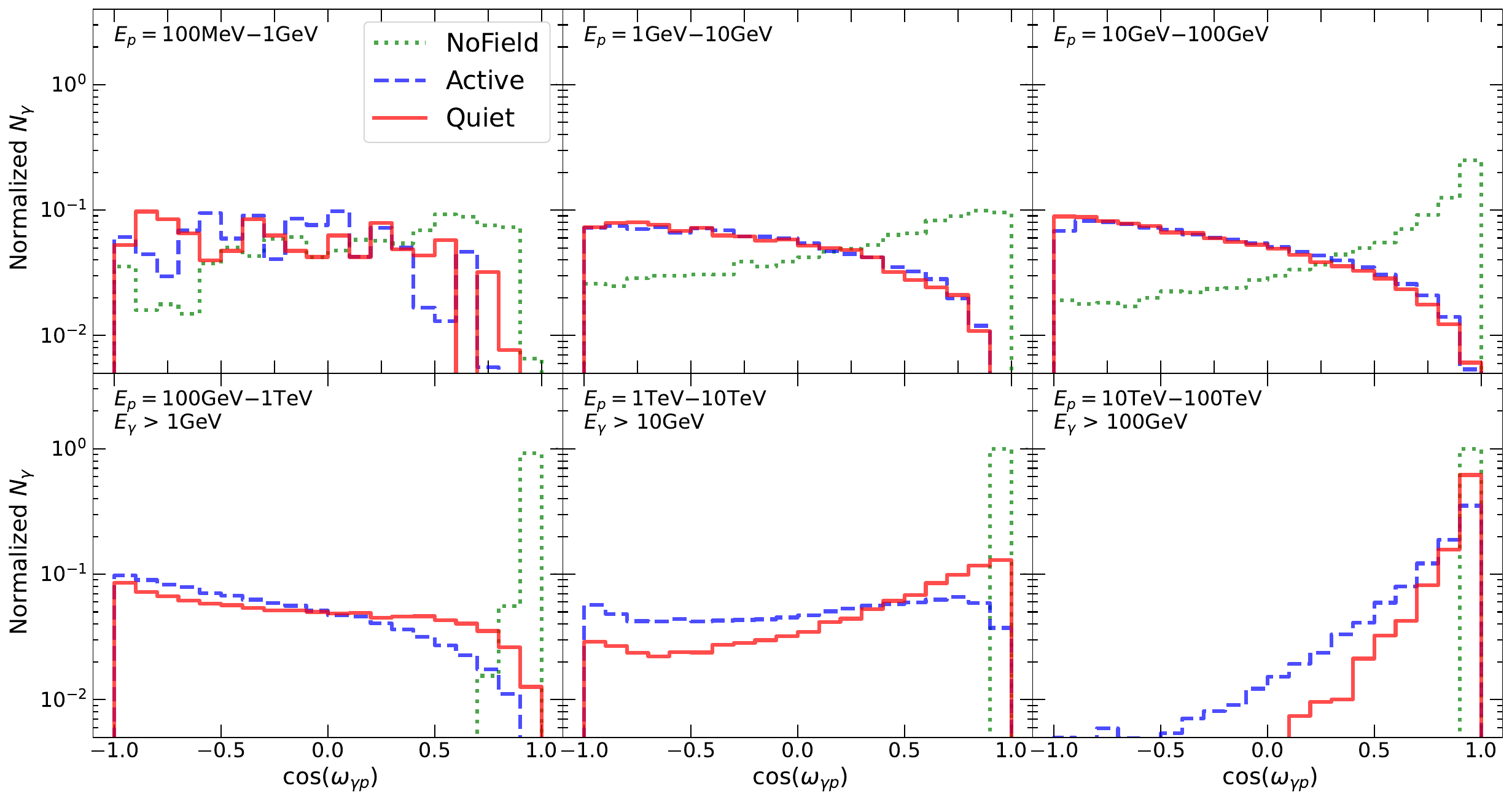}
    \caption{Similar to Fig.~\ref{fig:omega_gamma}, but for the distribution of the angle between each pair of incoming protons and outgoing gamma rays, $\omega_{\gamma p}$. At high energies, the peaked distributions corresponds to the outgoing gamma rays having nearly the same direction as the incoming protons, which reflects the high Lorentz factor effect. Taking Fig.~\ref{fig:omega_proton} and Fig.~\ref{fig:omega_gamma} into consideration reaffirms that this is the Sun-limb scenario. It is evident that the introduction of magnetic fields broaden this distribution, which enhances gamma-ray production.  In addition, the distribution for the \nof\ case is also broad at low energies, which is likely cased by kinematic effects. This is responsible for the enhanced \nof\ flux compared to the limb case. We also note that at the lowest energy bin, all three cases have similarly broad angular distributions, shows that the enhancement saturates, as shown in the lowest energy bins in Fig.~\ref{fig:flux_energy}.} 
    \label{fig:omega_gamma_proton}
\end{figure*}

\subsection{Physical Processes}\label{sec:production}

Figure~\ref{fig:production} shows the breakdown of the physical processes that contribute to the solar disk gamma-ray production. For all three cases, we find that the dominant contribution comes from neutral pion~($\pi^{0}$) decays.  These pions could be produced directly from the primary proton interactions, or from the subsequent hadronic showers. Due to the short lifetime of $\pi^{0}$, they decay promptly before undergoing additional scatterings. As a result, $\pi^{0}$ can efficiently convert the primary proton energy into gamma rays.  

The second most important source of gamma rays, at $\sim$\,10\% level, comes from electron and positron bremsstrahlung as well as positron annihilation~(labelled simply as $e^{+}e^{-}$).  These electrons and positrons come from the final states of many secondaries~(e.g., $\pi^{\pm}$), or they can be produced from electromagnetic showers initiated by energetic gamma rays or electrons.

Finally, we group the remaining gamma-ray production channels into ``others'', which includes decays of heavier hadron states (e.g., $\eta, \Sigma, \Omega$, etc), hadron inelastic scatterings, muon-bremsstrahlung, etc. These contributions are subdominant, but not negligible.

\subsection{Energy contribution}\label{sec:flux_energy}

Figure~\ref{fig:flux_energy} shows the contributions from each input proton energy intervals to the total gamma-ray flux. Interestingly, for proton energies 0.1\,GeV to 10\,GeV, we find no significant changes in the contribution between \nof\ and those with magnetic fields. Most of the flux enhancements for \quiet\ and \act\ come from proton energies from 10\,GeV to 1\,TeV, where the low energy ``tails'' in the \nof\ case change into ``bumps'' in the cases with magnetic fields.  From 1\,TeV to 100\,TeV, while the gamma-ray production is enhanced, the enhancements are mainly at low energies, which are buried by the other low-proton-energy components and have little effect to the final results.

\subsection{Event angular distribution}\label{sec:angular}

We explore in detail the angular distribution of protons that interact in the solar atmosphere as well as that of the outgoing gamma rays, which we find helpful in elucidating the the physics behind the enhanced gamma-ray production with magnetic fields. We consider three angles, $\omega_{p}$, $\omega_{\gamma}$, and $\omega_{\gamma p}$, which are illustrated schematically in Fig~\ref{fig:diagram}.

Figure~\ref{fig:omega_gamma} shows the angular distribution of the outgoing gamma rays. We define the angle $\omega_{\gamma}$ as the angle between the vector of the escaped gamma rays and the normal direction of the spherical simulation volume, evaluated at +3000\,km above the photosphere. In other words, $\cos(\omega_{\gamma})$ = 1, 0 correspond to gamma rays pointing radially outward and tangential to the simulation volume, respectively.

Similarly, Figure~\ref{fig:omega_proton} shows the angular distribution for the incoming protons, where the angle $\omega_{p}$ is defined by the angle between the proton vector and the normal direction at +3000\,km. We note that here we only consider protons that have successfully produced at least one escaped photon; protons that are completely absorbed or escaped without producing gamma rays are not included here. 

And finally, Figure~\ref{fig:omega_gamma_proton} shows the distribution for $\omega_{\gamma p}$, defined as the angles between the outgoing gamma rays and the incoming protons that produce gamma rays. Each proton could contribute multiple outgoing photons; all these pairs are considered in the distribution.

We note that for high proton energies, a large number of lower-energy gamma rays are produced, which are not important to the problem at hand, as shown in Sec.~\ref{sec:flux_energy}. Therefore, to highlight the angular distribution for the relevant photons, we apply gamma-ray energy cuts $E_{\gamma}>1$GeV, $E_{\gamma}>10$GeV, and $E_{\gamma}>100$GeV for the three proton energy intervals above 100\,GeV, respectively.

\subsubsection{Distribution without magnetic fields} \label{sec:nofield_angle}

It is instructive to first consider the \nof\ case. At high energies~($E_{p}>100$\,GeV), we find that the distributions tend towards $\omega_{\gamma}\simeq 90^{\circ}$, $\omega_{p}\simeq 90^{\circ}$, and $\omega_{\gamma p }\simeq 0^{\circ}$.  This peaked angular distribution appear naturally due to the large Lorentz factor~(except for the low-energy secondary photons that are cut from the figures). This corresponds to the Sun-limb scenarios, as discussed in Ref.~\cite{Zhou:2016ljf}.

However, for lower energy protons with energy less than roughly 100\,GeV, we find that the distribution become significantly broader. Because there are no magnetic fields to change the trajectories of the particles, the broader distribution must be caused by the scattering angular distribution itself. This is expected to be caused by large transverse momentum distribution for the pion produced after scattering~\cite{Karlsson:2007pt, Koers:2006dd}.

The broader angular distribution can also explain the difference between the calculations from Zhou et al.~\cite{Zhou:2016ljf} and the simulation results from this work and Gao et al.~\cite{Gao:2017bfv}. With the 1D approximation used in Zhou et al., gamma rays produced by protons with steep incident angles are all absorbed. 
However, as shown in our 3D calculation, e.g., the $\omega_{p}$ distribution in 1\,GeV-100\,GeV, the distribution is broad and even down-going protons ($\omega_{p}\sim 180^{\circ}$) can produce observable gamma rays. This is also precisely the proton energy range responsible for the enhanced gamma-ray flux~($E_{\gamma}\sim 1$\,GeV) in our simulation compared to that by Zhou et al. Therefore, we conclude that \emph{by kinematic effects alone, proton-proton scattering can produce large-angle events; and these large angle events are responsible for enhancing the gamma-ray production around 1\,GeV.} At higher energies, as gamma rays and protons become more collinear, the 3D and 1D calculations produce similar results.

\subsubsection{Distribution with magnetic fields}\label{sec:bfield_angle}

Comparing the angular distributions of \nof\ with that from \quiet\ and \act, we see that the PFSS magnetic fields have a significant impact on the distribution. This is expected as both the directions of the primary protons and the charged secondaries~($\pi^{\pm}$, $e^{\pm}$, etc) are bent as they propagate in the simulation volume. The bending effect is evidently shown in the $\omega_{p}$ distribution. Importantly, as the \nof\ distribution becomes more peaked for $E_{p}>100$\,GeV, the distribution with magnetic fields remain broad until $E_{p}>10$\,TeV. Following the observations from the previous section, the broader distribution is responsible for the enhanced gamma-ray production compared with the \nof\ case. We conclude that \emph{the broadening of the angular distributions, caused by magnetic fields bending the cosmic-ray primaries and secondaries, can enhance the solar gamma-ray production.} 

We note that our results with and without magnetic fields all have roughly the same flux at 100\,MeV. We believe this is somewhat accidental.  On one hand, the magnetic fields broaden the angular distribution and enhances gamma-ray flux production.  On the other hand, magnetic fields can also deflect the cosmic rays and reduce the overall production rate.  The latter effect may not be fully reflected in this calculation, as we consider consider a propagation space of 3000\,km above the photosphere. We leave a more thorough investigation of this for future works.

Finally, comparing between the \quiet\ and \act\ results, we find that the \act\ angular distribution has more large-angle contributions than that for \quiet. This also explains why our simulation results find that the \act\ flux is larger than the \quiet\ flux around 10\,GeV-1\,TeV. This follows from Tab.~\ref{table:PFSS}, where \act\ have higher typical magnetic fields than \quiet, thus leading to more magnetic bendings. On the other hand, the \quiet\ result is similar to the \act\ result around 1\,GeV. This suggests that below 1\,GeV, larger magnetic fields does not further enhance the gamma-ray production, which is also shown by the similar $\omega_{\gamma p}$ distribution for proton energy between 1\,GeV and 100\,GeV. In other words, the magnetic-field enhancement effect saturates.

\subsubsection{Estimating the Gyroradius}\label{sec:gyroradius}
It is instructive to estimate the gyroradius, which puts the length and energy scales into perspective. The shortest length scale of our simulation is in the radial direction, 3600\,km. Setting this as the gyroradius and consider 10\,G as a typical field strength~(Sec.~\ref{sec:PFSS}), the critical energy is 
\begin{equation}\label{eq:larmor}
    E_{c} \simeq {\rm 1\,TeV} \left( \frac{r}{3600\,{\rm km}} \right) \left( \frac{B}{10\,{\rm G}} \right) \, . 
\end{equation}
This means that it is possible for protons with energies below $E_{c}$ to undergo a complete reversal in their pointing direction in the simulation volume. Above $E_{c}$, one would then expect the effect of magnetic fields to decrease and approach the collinear limit. Indeed, this is consistent with observations from Figs.~\ref{fig:omega_proton} and~\ref{fig:omega_gamma_proton}, where we can see that the angular distributions undergo a qualitative change below and above TeV.

\subsection{Solar modulation} \label{sec:modulation}

\begin{figure}[t!]
    \centering
    \includegraphics[width=\columnwidth]{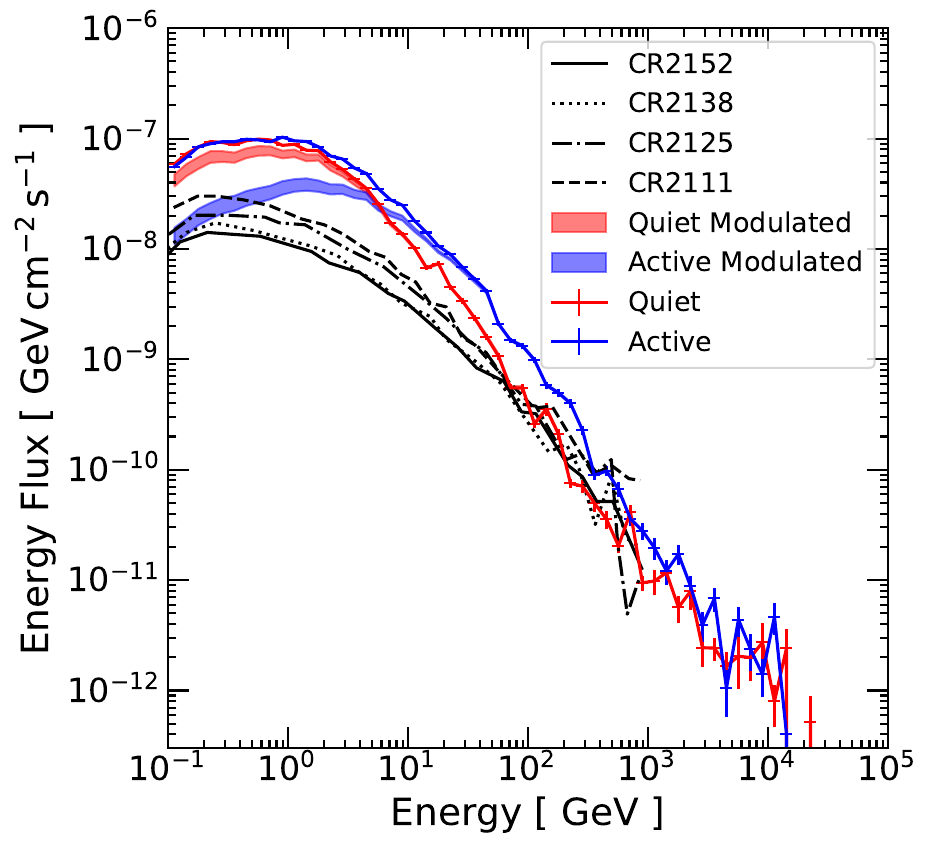}
    \caption{The bands show the effect of taking into account solar modulation on the solar gamma-ray production. The upper edge of the band corresponds to Model-2 from Ref.~\cite{Abdo:2011xn}, while the lower edge corresponds to Model-1. The force field potential for the modulated \quiet\ (\act) case is chosen to be 300 (1000)\,MV. For comparison, the un-modulated results from Fig.~\ref{fig:bfield_flux} are also shown. As shown, solar modulation could affect the final gamma-ray flux substantially at low energy, but negligibly for gamma rays above 10\,GeV. Simulation results including four magetific field models CR2152, CR2138, CR2125 and CR2111 from Ref.~\cite{Mazziotta:2020foa} also displayed. Detailed comparison is in Sec.\ref{sec:previous_work}} \label{fig:solar_mod}
\end{figure}

We have considered a fixed cosmic-ray spectrum measured at Earth for the incoming cosmic rays, ignoring the effect of solar modulation and additional modulation from Earth to the Sun. The latter, in particular, has substantial theoretical uncertainties.

To estimate the effect of solar modulation, we use the simple, yet effective force field approach~\cite{Gleeson:1968zza}. 
we follow Ref.~\cite{Abdo:2011xn} to compute the cosmic-ray spectrum near the solar surface.
The local interstellar spectrum, $F_{lis}(E)$, is taken from Ref.~\cite{2014SoPh..289..391P}. The cosmic-ray spectrum at different solar radii is then given by 
\begin{equation}
F_{p}(E,r)= F_{lis}(E) \frac{E^{2} - m_{p}^{2}c^{4}}{ \left(E+\Phi(r) \right)^{2} -m_{p}^{2}c^{4} }\, ,
\end{equation}
where $\Phi(r)$ is the modulation potential. We adopt the modulation potential that are first derived in Ref.~\cite{Moskalenko:2006ta}, based on the results from Ref.~\cite{fujii2005spatial}. 
Two modulation potentials are considered. Model-1, suitable for solar cycles 20/22, is 
\begin{equation}
\Phi_1(r)= \frac{\Phi_0}{1.88}\left\{
\begin{array}{ll}
r^{-0.4} - r_b^{-0.4}, & r\ge r_0,\\
0.24 + 8 (r^{-0.1} - r_0^{-0.1}), & r<r_0 
\end{array}
\right.
\end{equation}
and Model-2, suitable for solar cycle 21 is
\begin{equation}
\Phi_2(r)= \Phi_0 (r^{-0.1} - r_b^{-0.1}) / (1 - r_b^{-0.1}).
\end{equation}
Here $\Phi_0$ is the modulation potential at 1\,AU, $r_{0} = 10$\,AU, and $r_{b} = 100$\,AU. (We note that Model-3 in Ref.~\cite{Abdo:2011xn} assumes no additional solar modulation inside the Earth radius, which corresponds to our default assumption.)

Our default results (as in Fig.~\ref{fig:bfield_flux}) corresponds to having a force field potential at Earth position, $\Phi_{0}$, to be around 600\,MV. 
To estimate the effect of modulation, we consider $\Phi_{0}=300$\,MV and $\Phi_{0}=1000$\,MV for solar minimum and solar maximum, respectively.  Using these modulated spectra, we re-weigh our simulation result and obtain the correspond output gamma-ray flux. We note the models and the parameters chosen here do not correspond to the same periods as the measured Fermi data. Thus, this only served as a qualitative estimate on the effect of solar modulation. 

Figure.~\ref{fig:solar_mod} shows the modulated gamma-ray flux. The red~(blue) band shows the \quiet\ (\act) case, where the solar minimum (maximum) force field potential is used. The upper edge of the band corresponds to Model-2 and the lower edge corresponds to Model-1. 

As expected, the effect of solar modulation is mostly at low energies, with the suppression stronger at lower energies.  Due to the modulation, \act\ flux is now lower than the \quiet\ flux below a few GeV. Qualitatively, solar modulation effect could be responsible for some and possibly all of the time variation seen in the low-energy Fermi data~\cite{Tang:2018wqp, Linden:2020lvz}. We leave a detailed time-variation comparison with data for future works.

It is interesting to note that the effect of solar modulation diminishes above 10\,GeV.  Therefore, the large time variation seen above 100\,GeV~\cite{Tang:2018wqp, Linden:2018exo} cannot be solely explained by the cosmic-ray solar modulation models. 
This, yet again, provides evidence that other magnetic structures are needed to explain these events. It is also quite possible that they could affect the production of low-energy gamma-ray events. 

\subsection{Comparison with observations} \label{sec:observation}

\begin{figure*}[t!]
    \centering
    \includegraphics[width=180mm,scale=0.2]{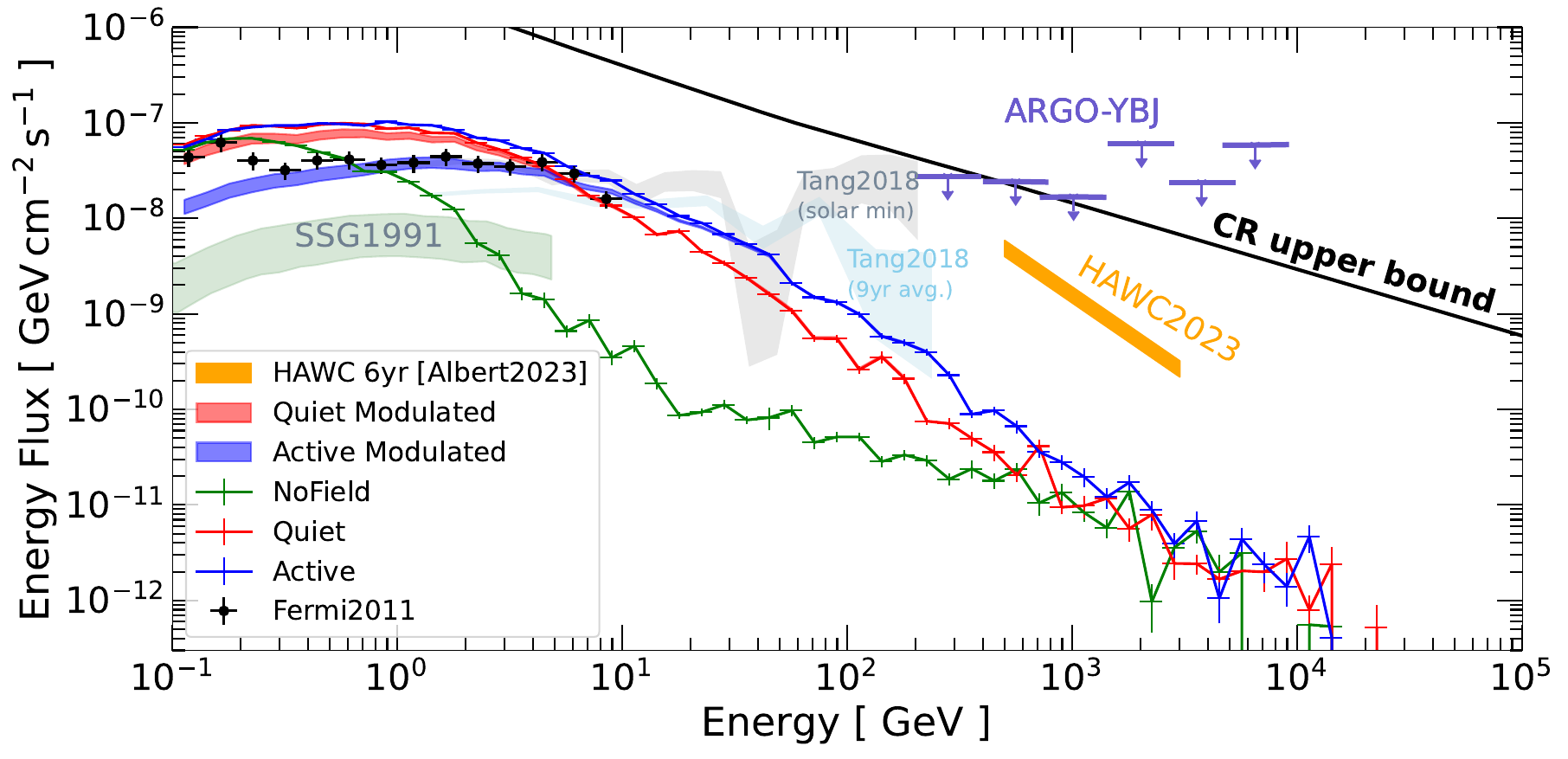}
    \caption{Our simulation results~(\nof, \quiet,\act, Quiet Modulated and Active Modulated) compared with solar disk observations from the 1.5-yr analysis by the Fermi collaboration~\cite{Abdo:2011xn}, follow-up analysis with a similar period by Tang et al.~\cite{Tang:2018wqp}, and their 9-yr averaged results. The comparison of the model with the Fermi-LAT data refers to different time intervals. The solar minimum case of Fermi-LAT is from 2008-8-7 to 2010-1-21, while the “Quiet” case in our model refers to a Solar magnetic configuration before the launch of the Fermi satellite from 2008-5-9 to 2008-6-9.  The simulated gamma-ray spectrum affected by solar modulation both for maximum and minimum activity models are also displayed. We also show the prediction from Seckel et al.~\cite{Seckel:1991ffa}~(SSG1991), and the theoretical upper bound from Linden et al.~\cite{Linden:2018exo}. The upper limits at high energies comes from ARGO-YBJ~\cite{Bartoli:2019xvu} and detection by HAWC~\cite{HAWC:2022khj}. }
    \label{fig:diskflux_all}
\end{figure*}

Figure~\ref{fig:diskflux_all} shows our simulated solar disk emission together with observational results and constraints. Results with solar modulation are shown in colored bands, and those without solar modulation are shown in solar lines with error bars correspond to uncertainties due to monte carlo simulations. The green solid line correspond to the case without considering any magnetic fields. For observation results, we show the $\simeq$\,1.5-year result~(Aug. 2008 -- Feb. 2010) obtained by the Fermi collaboration~\cite{Abdo:2011xn} and the $\simeq$\,1.5-year result~(Aug. 2008 -- Jan. 2010) obtained by Tang et al.~\cite{Tang:2018wqp} that extended the analysis to higher energy. During these periods, the Sun is dominantly in quiet states, which we simply denote as solar minimum.  We also show the 9-year analysis (Aug. 2008 -- Jul. 2017) by Tang et al. that covers both periods of low and high solar activities. 

At low energies, between 0.1\,GeV and 10\,GeV, our results with magnetic fields are comparable to the solar minimum flux up to a factor of a few. Interestingly, Fermi observation suggests that the flux at this energy range anti-correlates with the solar activities~\cite{Ng:2015gya, Tang:2018wqp}. This is consistent with our results with solar modulation. In short, PFSS magnetic fields with solar modulation could qualitatively explain the low-energy Fermi data.  However, as explained below, this picture may not be complete due to the discrepancy at high energies.

At higher energies, our results fall below the data quickly. At 100\,GeV, both the \quiet\ and \act\ results is about 1 order of magnitude less than the observation. In particular, the spectrum of solar minimum observation was found to be hard until at least 200\,GeV with no signs of cutoff, making the disagreement with our results even larger.  For the 9-year averaged spectrum, which is dominated by periods of higher solar activity, the spectrum softens rapidly above 100\,GeV, and could potentially fall into agreement with our calculations. 
The quality of measurement, however, is not sufficient to draw conclusive statements yet. 
Lastly, we also do not see the extreme time variability in the $>100$\,GeV photon flux and the spectral dip around 30-50\,GeV~\cite{Linden:2018exo, Tang:2018wqp}. Thus, new magnetic fields or new physics ingredients must be needed, in addition to coronal fields, to explain the high-energy photon flux. 
In order to produce 100\,GeV photons, roughly 1\,TeV cosmic rays are needed. Thesefore, the extreme time variability observed at 100\,GeV cannot be explained by modulation of cosmic rays alone. 

Above the Fermi-LAT energy range, only large ground-based air shower gamma-ray observatories can potentially detect high-energy gamma rays from the Sun. We show the upper limits from ARGO-YBJ~\cite{Li:2019ji} and the detection by HAWC~\cite{HAWC:2022khj}, both orders of magnitude higher than our calculation.  For reference, we also show the theoretical upper limit from cosmic rays interacting with the atmosphere~\cite{Linden:2018exo}.  Given that the solar minimum flux measurement by Fermi did not exhibit a cutoff, a detection could be possible with HAWC or LHAASO.  TeV detection or constraint will be essential for identifying the mechanism responsible for the high-energy flux. 

\subsection{Comparison with other calculations with magnetic fields} \label{sec:previous_work}

For many years, the only solar disk gamma-ray calculation that took into account magnetic fields was the pioneer work by Seckel, Stanev, and Gaisser~(SSG1991 \cite{Seckel:1991ffa}). In SSG1991, cosmic-ray propagation in the solar system was taken into account, and more importantly, cosmic rays entering the solar atmosphere were assumed to be funneled into magnetic flux tubes, and then reflected in the flux tubes due to the large field gradient. As a result, the gamma-ray production is enhanced by having the possibility of cosmic rays interacting after being reflected. However, even with such an enhancement, the SSG1991 model prediction is still much lower than the observation. Interestingly, in this work we find that at low energies, scattering kinematics and coronal magnetic fields can provide more than enough boost to the gamma-ray production. Thus, we find that the SSG1991 flux-tube reflection may be a subdominant mechanism for enhancing the gamma-ray production. However, flux tubes could still be important for bringing the 0.1--10\,GeV flux to quantitatively match the observational data. 

Compared to existing solar gamma-ray calculations with magnetic fields, Mazziotta et al.~\cite{Mazziotta:2020foa} have independently published a work that used another particle interaction simulation package \texttt{FLUKA}~\cite{Bohlen:2014buj} to simulate the solar disk gamma-ray production with PFSS magnetic fields. 
Compared to our results, Mazziotta et al. took into account cosmic-ray propagation in the solar system using a custom propagation code HelioProp instead of the force field model; they also employed a larger simulation volume filled with the PFSS magnetic fields. They consider the PFSS fields from 4 Carrington rotations from 2011 to 2014, roughly in the middle between a solar minimum and maximum. 
Despite all the differences, our results agree with Mazziotta et al. qualitatively in the sense that with PFSS magnetic fields, the $\sim 1$\,GeV gamma-ray production is boosted to close to the level of Fermi data.  Quantitatively, our results are a factor of a few higher than Mazziotta et al. when comparing more similar solar activity periods. This difference could be caused by differences in the chosen simulation volume, magnetic field resolution, and solar modulation models. Roughly, above 100\,GeV, our results reach good agreement with each other. 
Mazziotta et al. also showed that with an enhanced magnetic field profile near the photosphere~(the \texttt{BIFROST} profile), the gamma-ray production at higher energies can be further enhanced. This is in good agreement with our physical interpretation on the nature of the boost mechanism in Sec.\ref{sec:bfield_angle}.  However, with or without the boosted magnetic field profile, the Fermi gamma-ray data above 100\,GeV during solar minimum still cannot be explained. This agrees with our conclusion that coronal magnetic fields can not explain the high-energy disk emission. 

In Hudson et al.~\cite{Hudson:2020MNRAS}, the solar surface magnetic field configuration from inside the convection zone to the corona was overviewed in the context of charged particle propagation.  This include both cosmic rays as well as energetic particles accelerated by the Sun itself.  The discussion mostly focused on lower energy particles that follow the field lines closely, e.g., as in Ref.~\cite{Stecker:1973NPhS..242...59S,Ramaty:1975SSRv...18..341R}.  Nevertheless, the magnetic field structure discussion in this work will likely be an important addition in full calculation of solar particle propagation in the Sun. 

\subsection{Nuclei effects} \label{sec:nuclei}
We have only considered protons in both cosmic rays and in the solar atmosphere. Adding the nuclei species would slightly enhance the gamma-ray production. This is often taken into account by considering the so-called nuclear enhancement factor. Following Ref.~\cite{Kachelriess:2014mga}, we compute the enhancement factor due to p-He, He-p, and He-He interactions.  Considering the He/p number abundance ratio in the photosphere to be about 8\%~\cite{Grevesse:2002AdSpR} and the He/p ratio in the cosmic rays to be about 8-10\%, the combined enhancement factor due to Helium increase the gamma-ray flux production by about a factor of 1.6 to 1.8. The effect from other species are substantially smaller. 

The effect of nuclear species, though not exactly tiny, is subdominant to uncertainties associated with the complicated solar magnetic fields. Also, this small factor does not change our conclusions, we thus leave the full implementation of nuclear effects to future works.

\section{Conclusions and Outlook} \label{sec:conclu_and_outlook}

\subsection{Conclusion} \label{sec:conclusion}

In this work, we present our \geant\ based code \gsolar, which simulates cosmic-ray interactions in the solar atmosphere with magnetic fields. Using \gsolar, we compute the solar disk gamma-ray flux in three scenarios, without magnetic fields~(\nof), with coronal magnetic fields during low solar activity~(\quiet), and with magnetic fields during peak solar activity~(\act). We use the PFSS coronal magnetic field model, and we only focus in the volume from 600\,km below photosphere to 3000\, above, where interactions are expected to happen. It is clear that the inclusion of magnetic fields enhances the gamma-ray production compared to the \nof\ case~\cite{Hudson:2020MNRAS}. 

From the simulated gamma-ray flux spectrum and by studying their underlying composition and angular distributions, we have these main findings: 
\begin{itemize}
    \item Without magnetic fields, the solar disk flux production below 10\,GeV can be significantly enhanced due to photons produced with large scattering angles relative to the primary proton direction, caused purely by particle scattering kinematics. 
    \item With the PFSS magnetic fields, the solar disk flux production is further enhanced up to 1\,TeV. This is due to much wider angular distribution for the escaped gamma rays, caused by magnetic fields bending the trajectories of primary protons and charged secondaries.
\end{itemize}
While there are still significant quantitative disagreements between our result and the observations~\cite{Abdo:2011xn, Ng:2015gya, Linden:2018exo, Tang:2018wqp}, we believe this work has elucidated at least one pathway that could lead to a complete model for explaining the solar disk gamma-ray flux. 

\subsection{Outlook} \label{sec:outlook}

In this work, we find that coronal fields can not explain the observed gamma rays above 100\,GeV, especially during solar minimum. This suggests that features with stronger magnetic fields, e.g., sunspots or active regions, could be responsible for the production of high-energy gamma rays. However, this is contradictory to the observed time variation~\cite{Linden:2018exo, Tang:2018wqp}, where \emph{more} high-energy photons were observed from the Sun during solar minimum, when the number of sunspots are few. We leave these investigations for future works. 

One of the main simplifications in our simulation is taking the simulation space boundary to be 3000\.km above the photosphere. However, the effect of PFSS fields could affect the cosmic-ray propagation up to ${\cal O}(1)$ solar radii above the photosphere. Then at a distance further than that, solar modulation of cosmic rays should be the dominant effect. Thus, in our setup, the effect of the PFSS magnetic fields could be underestimated at low energies. We leave further investigation of this for future works.  

Our results between 0.1 to 10 GeV seems to agree with the observed data up to a factor of a few.  However, the data at about 100\,GeV differs from the calculation by roughly an order of magnitude, which suggests that additional physics inputs are needed to  qualitatively explain at high energies. Naturally, one would expect sources of strong magnetic fields responsible for the high-energy gamma rays could also affect the production of low-energy gamma rays. Therefore, without a model for the whole energy range, we believe it is too early to state that PFSS fields together with solar modulation are solely responsible for the $\sim$GeV solar gamma-ray observation. However, it is certain that PFSS fields plays an integral part at the production of $\sim$GeV solar gamma rays.

The gamma-ray data by Fermi-LAT~\cite{Abdo:2011xn, Ng:2015gya, Linden:2018exo, Tang:2018wqp} provide a rich set of phenomena that is not touched nor explained in this work, including time variations, spectral features, and gamma-ray morphology.  Furthermore, cosmic-ray Sun shadows~\cite{Amenomori:2013own,2018PhRvL.120c1101A, Aartsen:2018zrl, BeckerTjus:2019rqu, Aartsen:2020hzn} should be intimately related to the production of the disk gamma rays~\cite{Gutierrez:2019fna}. We anticipate that once the relevant magnetic fields are identified or included in the calculation, these features and observations will be important for verifying or differentiating competing models. 

In the very-high-energy regime, HAWC and LHAASO could provide valuable information for understanding the production of solar gamma rays.  Recent results by HAWC~\cite{HAWC:2022khj} have shown that the Sun continues to emit gamma rays at TeV energy range, but with a much softer spectrum.  LHAASO, with a larger collecting area, is expected to produce more precise measurement. It is clear from this work (also Ref.~\cite{Mazziotta:2020foa}) that PFSS magnetic fields could not account the TeV solar emission.  New magnetic-field structures or ideas, e.g., Ref~\cite{Banik:2023shc}, is needed to explain these TeV solar gamma-ray flux.


High-energy neutrinos are inevitably produced together with the solar disk gamma-ray flux~\cite{Moskalenko:1991hm, Seckel:1991ffa, Moskalenko:1993ke, Ingelman:1996mj, Arguelles:2017eao, Edsjo:2017kjk, Masip:2017gvw, Mazziotta:2020foa}, and could potentially be detected by neutrino telescopes~\cite{Aartsen:2019avh}. At the same time, the Sun is an important target for dark matter searches, where the signal could be anamalous neutrinos~\cite{Press:1985ug, Krauss:1985ks, Silk:1985ax, Gould:1987ir, Gould:1987ju, Peter:2009mk, Bell:2011sn, Niblaeus:2019gjk} and gamma rays~\cite{Batell:2009zp, Leane:2017vag, Arina:2017sng}.  Searches of these signals have yielded some of the strongest dark matter constraints in the literature~\cite{Aartsen:2016zhm, Adrian-Martinez:2016gti, Adrian-Martinez:2016ujo,  Albert:2018jwh,Cuoco:2019mlb, Mazziotta:2020foa}.  Having a robust model of cosmic rays interacting with the Sun is important for getting an accurate background estimate for these dark matter searches~\cite{Arguelles:2017eao,Ng:2017aur, Edsjo:2017kjk}. 

Ultimately, a precise understanding of how cosmic rays interact with the Sun could have the potential of allowing high-energy gamma rays and neutrinos as new windows for probing solar magnetic fields~\cite{Strauss:2012zza,Owens2013,Potgieter:2013pdj,Solanki:2010je,Wiegelmann:2012mu,Mackay:2012ww,Yang:2020science}, and could offer new perspectives in solar physics.

\section*{\label{sec:acknowledgements} Acknowledgments}

We thank John Beacom, Annika Peter, Tim Linden, Mehr Un Nisa, Bei Zhou, and Guanying Zhu for helpful comments. We also acknowledge the essential support of Guofu Cao, Yiqing Guo, Shuangquan Liu, and Cong Li, in the developing of the \gsolar\ code. This work is supported in China by NSFC (No.12261160362, No.12022502). KCYN is supported by grants provided by NSFC~(No.12322517, No.N\underline{~}CUHK456/22) and RGC (No.24302721, No.14305822, and No.14308023).

\bibliography{bib.bib}

\end{document}